\documentclass{rjparticle}
\usepackage{graphicx}

\newcommand{\miktex}{\hbox{Mik\kern-.15em\TeX}}

\title{Footprints of Higher-Dimensional\\ Decaying Black Holes}
\author[1]{Panagiota Kanti}
\affil[1]{Division of Theoretical Physics, Department of Physics, \\
University of Ioannina, Ioannina GR-45110, Greece\\ Email:{\em pkanti@cc.uoi.gr}}

\begin{document}
\maketitle
\begin{abstract}
We review the current results for the emission of Hawking radiation by a
higher-dimensional black hole during the Schwarzschild and the spin-down
phases. We discuss particularly the role of the angular variation of the
emitted radiation on the brane during the latter phase, the radiation spectra
for gravitons in the bulk, and the effect of the mass of the emitted particles
in determining  the bulk-to-brane energy balance.
\end{abstract}

\section{Introduction}

The new theories \cite{kanADD, kanAADD, kanRS} postulating the existence of
additional spacelike dimensions in nature implemented also a new fundamental
scale of gravity which could be considerably lower than the 4-dimensional one.
This quickly gave rise to the idea of the manifestation of strong-gravity
phenomena at high-energy particle collisions, among them the potential creation
of black holes \cite{kanBF, kanDL, kanGT}. 

The properties of these miniature black holes have been exhaustively studied
\cite{kanreview, kanreview2} including the emission of Hawking radiation which
will be the main direct observable effect associated with them. The latter
effect is expected to take place during the two intermediate phases in the
life of the black hole, the spin-down and the Schwarzschild phase \cite{kanGT}.
Whereas the study of the spherically-symmetric Schwarzschild phase is now
complete with the radiation spectra having all been derived and the question
of the bulk-to-brane energy balance been addressed, these same questions
remain open in the case of the axially-symmetric spin-down phase. In the
case of the emission along our brane, the radiation spectra for all Standard
Model particles have been indeed derived, nevertheless, these cannot currently
lead to the determination of both the angular-momentum of the black hole and
the number of extra dimensions as was initially hoped. In addition, the
derivation of the radiation spectra for the emission of gravitons in the bulk
is still in progress, and the question of whether the black hole emits most
of its energy in the bulk or on the brane during its spin-down phase remains
unanswered.  

In this talk, we will review what is known for the emission of Hawking
radiation on the brane by a black hole during the Schwarzschild and the spin-down
phases - we will particularly address the problem of extracting information on
the topological parameters of the theory from the brane radiation spectra, and
propose a solution related to the angular variation of the emitted radiation.
Then, we will discuss the question of the bulk-to-brane energy balance, the
role of the emission of gravitons and of the mass of the emitted particles in
determining the dominant channel.


\section{Creation and Properties of Higher-Dimensional Black Holes}

Throughout this talk, we will adopt the scenario of Large Extra Dimensions
\cite{kanADD} according to which a number $n$ of additional, flat, spacelike
dimensions exist in nature apart from the usual three. For simplicity,
these extra dimensions are assumed to have the same size ${\cal R}$
and to make up a compact space of volume $V \sim {\cal R}^n$.
Our $(3+1)$-dimensional world
is a hypersurface, a {\it 3-brane}, embedded in the $(4+n)$-dimensional
spacetime, the {\it bulk}. On our brane the usual Standard-Model particle
physics holds, with the gravitational interactions becoming strong only
at the Planck scale $M_P=10^{19}$\,{\rm GeV}. Unlike the 
Standard-Model particles that are localised on the brane, gravitons
are allowed to propagate in the whole of spacetime and mediate a 
higher-dimensional gravitational force -- this means that if two test
particles with masses $m_1$ and $m_2$ are brought at a distance $r \ll {\cal R}$,
the corresponding force will vary with $r$ as $1/r^{n+2}$. Moreover, one
may further assume that the magnitude of the gravitational force will be
determined by a higher-dimensional Newton's constant $G_D$; demanding that,
in the limit where $r$ increases at values much larger than ${\cal R}$,
contact should be made with the 4-dimensional expression,
it is found that \cite{kanADD}
\begin{equation}
G_D \simeq G_4\,{\cal R}^n\,.
\label{Newton's}
\end{equation}
If we further define $G_D \sim 1/M_*^{n+2}$, and use the relation
$G_4 \sim 1/M_P^2$, the above formula takes the form
\begin{equation}
M_P^2 \simeq {\cal R}^n \,M_{*}^{2+n}\,,
\label{Planck}
\end{equation}
where $M_*$ is the energy scale where gravitational forces become
strong in the context of the higher-dimensional theory. If, unlike string theory,
we assume that the size of the extra compact dimensions is much larger than
the Planck length, then $M_*$ can be considerably smaller than $M_P$. 

The most optimistic scenario assumes that $M_*$ can be as low as a few TeV,
an energy scale that is accessible at current particle-physics collision experiments.
Therefore, if the center-of-mass energy $E$ exceeds $M_*$, then, strong gravity
phenomena may become manifest. One of these phenomena is the creation of
a black hole \cite{kanBF} during the collision of two ordinary brane-localised
particles with impact parameter $b$: if $b<r_H(E)$, where $r_H(E)$ is the
Schwarzschild radius that corresponds to $E$, then a black hole will be formed
according to the Hoop Conjecture \cite{kanThorne, kanNewHoop}.  

The produced black hole, being a gravitational object, will be generically 
higher-dimensional and will extend both along and off the brane. In the 
limit $r_H \ll {\cal R}$, the black hole lives in a
$(4+n)$-dimensional spacetime where no distinction can be made between the
$n$ compact spacelike dimensions and the three infinite-sized ones. 
The simplest prototype that we may adopt to describe such a higher-dimensional
black hole is the spherically-symmetric Schwarzschild-Tangherlini one
\cite{kanTangherlini, kanMP}
\begin{equation}
ds^2 = - \left[1-\left(\frac{r_H}{r}\right)^{n+1}\right]\,dt^2 +
\left[1-\left(\frac{r_H}{r}\right)^{n+1}\right]^{-1}\,dr^2 + r^2 d\Omega_{2+n}^2\,,
\label{kanST}
\end{equation}
where $d\Omega_{2+n}^2$ is the line-element of a $(2+n)$-dimensional unit sphere
\begin{equation}
d\Omega_{2+n}^2=d\theta^2_{n+1} + \sin^2\theta_{n+1} \,\biggl(d\theta_n^2 +
\sin^2\theta_n\,\Bigl(\,... + \sin^2\theta_2\,(d\theta_1^2 + \sin^2 \theta_1
\,d\varphi^2)\,...\,\Bigr)\biggr)\,.
\label{kanunit}
\end{equation}
If we apply the Gauss law in $D=4+n$ dimensions, we find the following relation
between the black-hole horizon radius and its mass  \cite{kanMP}
\begin{equation}
r_H = {1\over M_*} \left(M_{BH}\over M_*\right)^{1\over n+1}
\left(8 \Gamma(\frac{n+3}{2}) \over (n+2) \sqrt{\pi}^{(n+1)}\right)^{1/(n+1)}\,.
\label{kanhorizon}
\end{equation}
The above formula is a generalised one holding for a higher-dimensional black
hole -- the well-known linear relation between the mass and the horizon radius
holding in 4 dimensions arises if we set $n=0$ in the above expression, and also
replace the fundamental Planck scale $M_*$ with the 4-dimensional one $M_P$. 

Although a Quantum Theory of gravity would be the natural framework in which
such high-energy particle collisions and their strong-gravity consequences
should be studied, one may draw valuable conclusions by using purely classical
arguments. Therefore, one may assert that a black hole could be created if the
Compton wavelength $\lambda_C=4\pi/E$ of the colliding particle of energy $E/2$
lies within the corresponding Schwarzschild radius $r_H(E)$ \cite{kanMR}.
By using the above-derived expression for the horizon radius (\ref{kanhorizon}),
we may write this creation condition in the form
\begin{equation}
\frac{4\pi}{E} < {1\over M_*} \left(E\over
M_*\right)^{1\over n+1} \left(8 \Gamma(\frac{n+3}{2}) \over (n+2)
\sqrt{\pi}^{(n+1)}\right)^{1/(n+1)}\,.
\label{kancriterion}
\end{equation}
The above formula may be solved for the ratio $x_{min}=E/M_*$ that determines the value
of the center-of-mass energy of the experiment necessary for the creation of the
black hole. We find that, for $n=2-7$, the minimum energy of the
collision should be $E=(8.0-11.2)\,M_*$, respectively \cite{kanMR}. As the 
maximum center-of-mass energy that will be achieved at the Large Hadron Collider
at CERN is 14 TeV, we immediately see that a window of energy of maximum width
of 6 TeV, in the optimistic case where $M*=1\,$TeV, will be open for the
search of miniature black holes.

In order for the produced black hole to be considered as a classical black hole,
its mass should be at least a few times larger than the fundamental scale of
gravity. As an indicative case, we may consider $M_{BH}=5$\,TeV for $M_*=1$\,TeV.
Then, Eq. (\ref{kanhorizon}) can give us the value of the horizon radius: as
$n$ varies from 1 to 7, we obtain $r_H=(4.06-1.99)\,10^{-4}$\,fm \cite{kanreview}.
Note that the presence of the new scale of gravity $M_*$ in Eq. (\ref{kanhorizon}) 
ensures that the horizon radius increases by more than 30 orders of magnitude
compared to its four-dimensional value for the same black-hole mass. 

A property of the black hole, that may be associated with observable
signatures of their creation, is its temperature. This is defined in terms
of the surface gravity of the black hole, and for the spherically-symmetric
case of Eq. (\ref{kanST}) is given by 
\begin{equation}
T_{H} = \frac{k}{2\pi}={(n+1) \over 4\pi\,r_H}\,.
\label{kantemp}
\end{equation}
By using the above formula and the corresponding values of $r_H$, we find that
the temperature ranges from 77 to 629 GeV for $n=1-7$, respectively \cite{kanreview}.
The value of the temperature defines the peak of the radiation spectra
associated with the Hawking process, i.e. the emission of elementary particles
by the black hole, and that clearly lies in a regime accessible by present-day
detection techniques.

As is well-known, no particle can escape through the horizon of the black hole.
Nevertheless, when a virtual pair of particles is produced outside the horizon
of the black hole and the antiparticle falls inside the black hole, then the
particle is free to propagate towards infinity. The particles that reach
asymptotic infinity make up the Hawking radiation which is characterized
by a thermal spectrum \cite{kanHawking, kanUnruh}
\begin{equation}
\frac{\textstyle dE(\omega)}
{\textstyle dt} = \int\,\frac{\textstyle {  |{\cal A}(\omega)|^2} \,\,\omega}
{\textstyle \exp\left(\omega/T_{H}\right) \mp 1}\,\,
\frac{\textstyle d\omega}{\textstyle (2\pi)}\,.
\label{kanrateS}
\end{equation}
In the above, $\omega$ is the energy of the emitted particles and $\pm 1$ a
statistics factor for fermions and bosons, respectively. The factor 
${|{\cal A}(\omega)|^2}$ is the Absorption  Probability (or, {\it greybody
factor}) and determines the number of particles that will escape the strong
gravitational field of the black hole to reach infinity.   

If we assume that the evolution of a higher-dimensional black hole is similar
to the one of its 4-dimensional analog, we expect the emission of Hawking radiation
to be realised during the two intermediate stages of the life of a black hole,
the axially-symmetric {\it spin-down} phase and the spherically-symmetric
{\it Schwarzschild} one \cite{kanGT}. The spin-down phase emerges after the
{\it balding} phase, during which the black hole forms and sheds all quantum
and classical conserved charges apart from the ones dictated by the no-hair
theorem of General Relativity. The {\it Planck} phase comes after the emission
of Hawking radiation has reduced the black-hole mass at the level of $M_*$:
this quantum object may either evaporate completely, having a lifetime of
the order of $\tau=10^{-26}$ sec \cite{kanADMR}, or reduce to a quantum remnant. 


\section{Decay of Higher-Dimensional Black Holes on the Brane}

The simplest phase in the life of a black hole is the one when it is characterised
by spherical symmetry. This phase follows when the black hole has emitted all, or
almost all, of its angular momentum. The gravitational background around such a
black hole is given by Eq. (\ref{kanST}). Since we are ourselves observers 
restricted to live on the brane, we are primarily interested in the emission of
particles by the black hole on our brane. To
study this effect, we need to derive the equation of motion of particles
with arbitrary spin in the gravitational background that is induced by the
black hole on the brane. The latter is found by setting the values of all 
additional $\theta_i$ coordinates, with $i=2,...,n+1$, to $\frac{\pi}{2}$ in
Eq. (\ref{kanST}). Then, the brane line-element takes the form
\begin{equation}
ds^2_4 = - \left[1-\left(\frac{r_H}{r}\right)^{n+1}\right] dt^2 +
\left[1-\left(\frac{r_H}{r}\right)^{n+1}\right]^{-1} dr^2 + 
r^2\,d\Omega_2^2\,.
\label{kanprojectedSchw}
\end{equation}

The equations of motion of Standard-Model-like fields with spin $s=0,1/2,1$ can
be put in the form of a ``master'' equation \cite{kanTeukolsky, kanreview, kanKMR}
by using the Newman-Penrose method \cite{kanNP, kanChandra}
and a factorized ansatz for the wavefunction of the field
\begin{equation}
\Psi_s=e^{-i\omega t}\,e^{im\varphi}\,\Delta^{-s}\,R_s(r)\,S_{s l}^m(\theta)\,.
\end{equation}
Then, the general field equation reduces to two decoupled equations, one for the
radial function $R_s(r)$ and one for the spin-weighted spherical harmonics
$S_{s l}^m(\theta)$ \cite{kanGoldberg}. The radial equation has the form
\begin{equation}
\Delta^s \frac{\textstyle d}
{\textstyle dr}\left(\Delta^{1-s}\frac{\textstyle d R_s^{~}}
{\textstyle dr}\right)+ \left[\frac{\textstyle \omega^2 r^2}{\textstyle h}+2i\omega s r-
\frac{\textstyle is\omega r^2 h'}{\textstyle h}
-\lambda_{s l}\right]R_s (r)=0\,,
\end{equation}
where $\Delta\equiv r^2\,h
\equiv r^2\left[1- \left(\frac{r_H}{r}\right)^{n+1}\right]$ -- note that,
through the non-trivial $n$-dependence of the metric function, the radial equation
of motion of a brane-localised field will  depend on the
number of transverse-to-the-brane extra spacelike dimensions.

The angular part of the master equation has the well-known form of the
differential equation satisfied by the spin-weighted spherical harmonics
\begin{equation}
\frac{1}{\sin\theta}\,\frac{\textstyle d}
{\textstyle d\theta}\left(\sin\theta\frac{\textstyle d S^m_{s l}}
{\textstyle d\theta}\right)+ \left[-\frac{\textstyle 2ms\cot\theta}{\textstyle \sin\theta}
-\frac{m^2}{\sin^2\theta}+s-s^2\cot^2\theta+\lambda_{s l}\right]S^m_{s l}(\theta)=0\,,
\end{equation}
where $\lambda_{s l}=l(l+1)-s(s-1)$. Due to the
spherical symmetry of the gravitational background, the emitted radiation from
the black hole will be evenly distributed over a $4\pi$ solid angle, and thus
the angular equation offers no new information.

\begin{figure}[t!]
\centering
\includegraphics[width=0.55\textwidth]{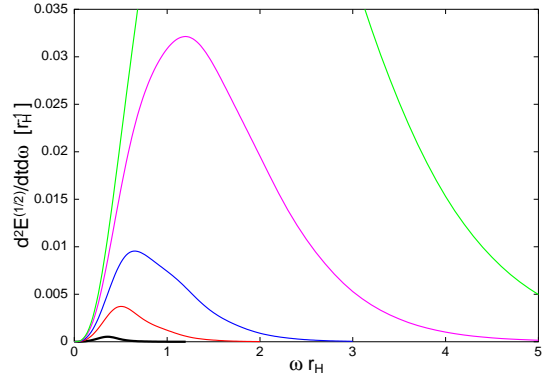}
\caption{Energy emission rates for fermions on the
brane or $n=0,1,2,4$ and 6 (from bottom to top) \cite{kanreview}.}
\label{kanfermions}
\end{figure}

We thus turn our attention to the radial part of the equation, and observe
that its solution for the radial function $R_s(r)$ determines the Absorption
Probability through the formula
\begin{equation}
|{\cal A}(\omega)|^2 \equiv 1-|{\cal R}(\omega)|^2 \equiv \frac{\textstyle 
\hspace*{2mm}{\cal F}_{\rm horizon_{~}}} 
{\textstyle {\cal F}^{~}_{\rm infinity}}\,,
\end{equation}
where ${\cal R}(\omega)$ is the Reflection coefficient and ${\cal F}$ the flux
of energy towards the black hole. The task of solving the radial equation
has been accomplished in the literature
both analytically \cite{kanKMR, kanIOP1} and numerically \cite{kanHK}.
By studying the scattering problem in the aforementioned black-hole
background, one may determine the value of the greybody factor, which when substituted
in Eq. (\ref{kanrateS}), can provide the radiation spectra of the emission of
different species of fields on the brane. In Fig. \ref{kanfermions}, we depict
the differential emission rate per unit time and frequency for fermions, in
terms of the number of transverse spacelike dimensions $n$. The energy emission
rate for fermions on the brane is greatly enhanced by the number of additional
spacelike dimensions, and this holds also for scalar and gauge bosons as may
be seen from the entries of Table \ref{kantableS}, where the total emissivities
have been normalised to the ones for $D=4$ \cite{kanHK}.

\begin{table}[b!]
\caption{Total emissivities for brane-localised scalars, fermions and gauge
bosons \cite{kanHK}}
\begin{center}
\begin{tabular}{|c|cccccccc|} \hline
$n$ & 0  & 1 & 2 & 3 & 4 & 5 & 6 & 7 \\ \hline
 Scalars  & 1.0 & 8.94 & 36.0 & 99.8 & 222 &
429 & 749 & 1220\\ 
 Fermions  & 1.0 & 14.2 & 59.5 & 162 & 352 &
664 & 1140 & 1830\\ 
G. Bosons  & 1.0 & 27.1 & 144 & 441 & 1020 &
2000 & 3530 & 5740 \\ \hline
\end{tabular}
\label{kantableS}
\end{center}
\end{table}

The situation in the case of the emission of Hawking radiation by a 
spherically-symmetric black hole is thus particularly simple. For a given
mass, the only other parameter characterising the gravitational background
is the number of additional spacelike dimensions $n$.
Thus, comparing the predicted emission rates with the observed ones, one may
even hope to determine the value of $n$. The same number determines also
the species of particles preferably emitted by a black hole: lower ($n=0,1$)
and higher-dimensional ($n=5,6$) black holes prefer to emit scalars
and gauge bosons, respectively, while black holes with $n=2,3,4$ have  a
more `democratic' type of spectrum \cite{kanHK}.

In the case, however, of a higher-dimensional rotating black hole, that emits
Hawking radiation, the situation changes considerably. The gravitational
background around such a black hole is significantly more complicated and
its line-element takes the form of the Myers-Perry solution \cite{kanMP}
\begin{eqnarray}
\hspace*{-1cm}
ds^2 &=& \left(1-\frac{\mu}{\Sigma\,r^{n-1}}\right)dt^2+\frac{2 a\mu\sin^2\theta}
{\Sigma\,r^{n-1}}\,dt\,d\varphi-\frac{\Sigma}{\Delta}dr^2 \nonumber \\[1mm] 
&-& \hspace*{0.3cm}
\Sigma\,d\theta^2-\left(r^2+a^2+\frac{a^2\mu\sin^2\theta}{\Sigma\,r^{n-1}}\right)
\sin^2\theta\,d\varphi^2 - r^2 \cos^2\theta\,d\Omega^2_n\,,
\label{kanMPmetric}
\end{eqnarray}
where now
\begin{equation}
\Delta=r^2+a^2-\frac{\mu}{r^{n-1}}\,, \quad \quad\Sigma=r^2+a^2\cos^2\theta\,.
\end{equation}
The above line-element describes the particular case of a `simply-rotating'
black hole, i.e. a black hole with only one non-vanishing angular momentum
component -- this is justified by the fact that the black hole has been
created by particles that are localised on the brane and have a non-zero impact
parameter only along a brane spacelike coordinate. The parameters $\mu$ and $a$
are then associated to the black hole mass and angular momentum, respectively,
through the relations
\begin{equation}
M_{BH}=\frac{(n+2) A_{2+n}}{16 \pi G}\,\mu \qquad {\rm and} \qquad 
J=\frac{2}{n+2}\,a\,M_{BH}\,, \label{kan-mu}
\end{equation}
where $A_{2+n}$ is the area of a $(2+n)$-dimensional unit sphere. The horizon
radius is found by setting $\Delta(r_H)=0$ and is found to be:
$r_H^{n+1}=\mu/(1+a_*^2)$, where we have defined the quantity $a_* \equiv a/r_H$.

We are again primarily interested in the radiation spectra emitted by the
rotating black hole on the brane. The line-element that is seen by the
brane-localised Standard Model fields follows again by fixing the values of
the ``extra'' angular coordinates -- in that case, the $d\Omega_{n}^{2}$ part
of the metric (\ref{kanMPmetric}) disappears while the remaining stays unaltered. 
A similar analysis, as in the case of a spherically-symmetric black hole, leads
again to a master equation for the propagation of an arbitrary spin-$s$ field
on the brane background. For the particular line-element, this general equation
decouples to a radial and angular part, given respectively by the following
equations \cite{kanreview, kanCKW}
\begin{equation}
\Delta ^{-s}\,\frac {d}{dr} \left(\Delta^{s+1}\,
\frac {d R_s}{dr}\right) + \left[\frac{K^2-iKs\Delta '}{\Delta}+4i s\omega r +
s\left( \Delta '' -2 \right)\delta_{s,|s|}-\Lambda^m_{sj}\right] R_s=0 
\label{kan-radialrot}
\end{equation}
and
\begin{eqnarray} && \hspace*{-1.0cm}\frac{1}{\sin\theta}\,\frac{\textstyle d}
{\textstyle d\theta}\left(\sin\theta\frac{\textstyle d S^m_{sj}}
{\textstyle d\theta}\right)+ \left[-\frac{\textstyle 2ms\cot\theta}{\textstyle \sin\theta}
-\frac{m^2}{\sin^2\theta}+a^2\omega^2\cos^2\theta\right. \nonumber \\[4mm]
&&\hspace*{3cm} \left.- 2 a \omega s \cos\theta +
s-s^2\cot^2\theta+\lambda_{sj}\right]S^m_{sj}(\theta)=0\,,
\label{kan-angulareq}
\end{eqnarray}
where we have used the definitions
\begin{equation}
K=(r^2+a^2)\,\omega-am\,, \qquad \Lambda^m_{sj}=\lambda_{sj} + a^2 \omega^2
-2 am \omega\,.
\end{equation}
The angular eigenvalue $\lambda_{sj}$ appearing in the angular equation does
not exist in closed form. It may be computed either analytically, through a power series
expansion in terms of $a \omega$ \cite{kanStaro, kanFackerell, kanSeidel} or numerically
\cite{kanCKW, kanHK2, kanDHKW, kanCDKW}. 

As in the case of spherically-symmetric black holes, one may compute the
emission spectra for all species of particles propagating in the gravitational
background of the rotating black hole on the brane. For this, we need the
value of the Absorption Probability that can be found by solving the radial
equation either analytically \cite{kanCEKT2, kanCEKT3} or numerically
\cite{kanCKW, kanHK2, kanDHKW, kanCDKW, kanFS, kanIOP2, kanIOP3}. Then, the
differential energy emission rate is given by the formula 
\cite{kanHawking, kanUnruh, kanOW}
\begin{equation}
\frac{d^2\,E}{\,dt\,d\omega}= \frac{1}{2\pi}\sum_{j,m}\,
\frac{{  |{\cal A}(\omega)|^2}\,\omega}{\exp(\tilde\omega/T_H)\mp 1}\,, 
\end{equation}
while the temperature and rotation velocity of the black hole are given by
\begin{equation}
T_{H}=\frac{(n+1)+(n-1)\,a_*^2}{4\pi(1+a_*^2)\,r_{H}}\,,
\qquad \Omega_{H}=\frac{a}{(r_H^2+a^2)}\,, \label{kan-Temprot}
\end{equation}
and $\tilde \omega= \omega -m \Omega_H$. 

\begin{figure}[t!]
\begin{center}
\mbox{ 
\includegraphics[scale=0.5]{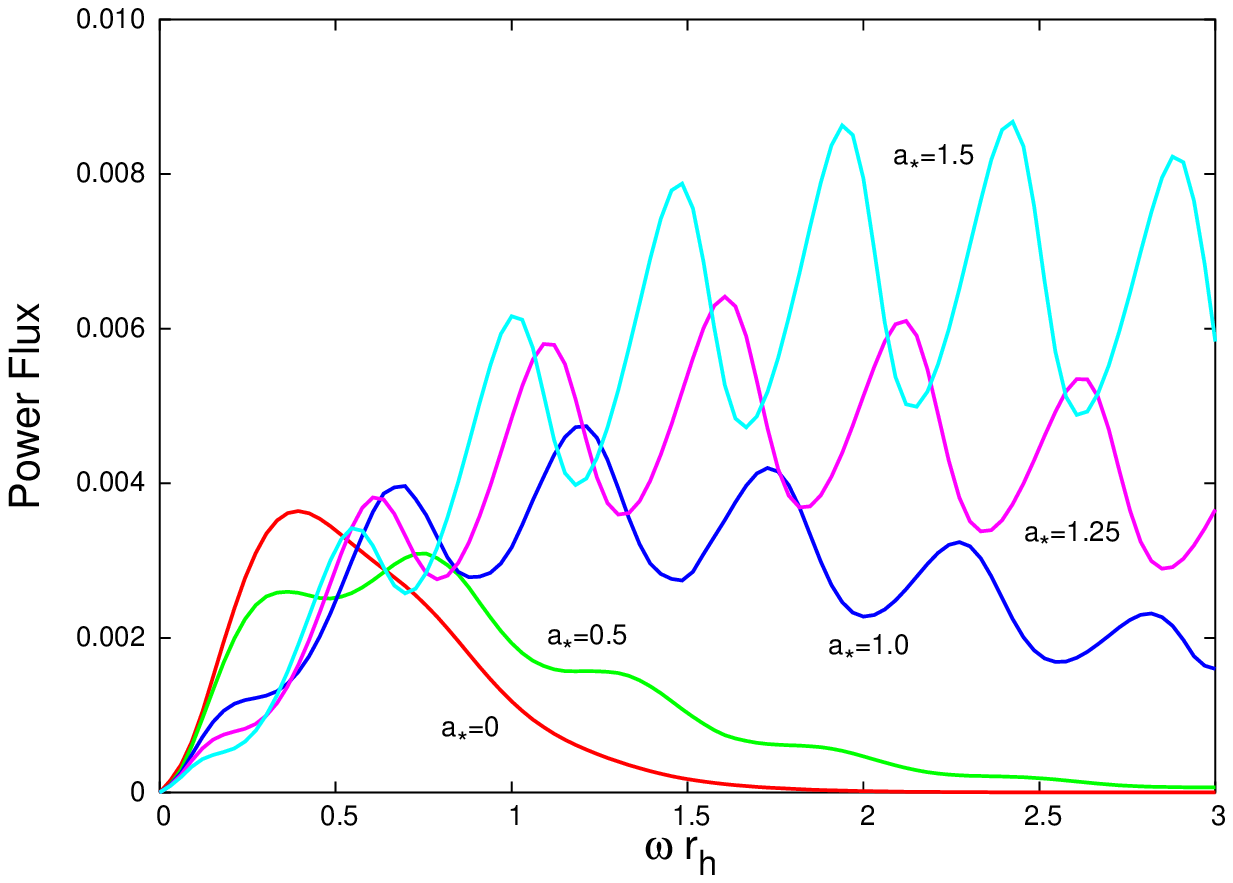}}
{\includegraphics[scale=0.5]{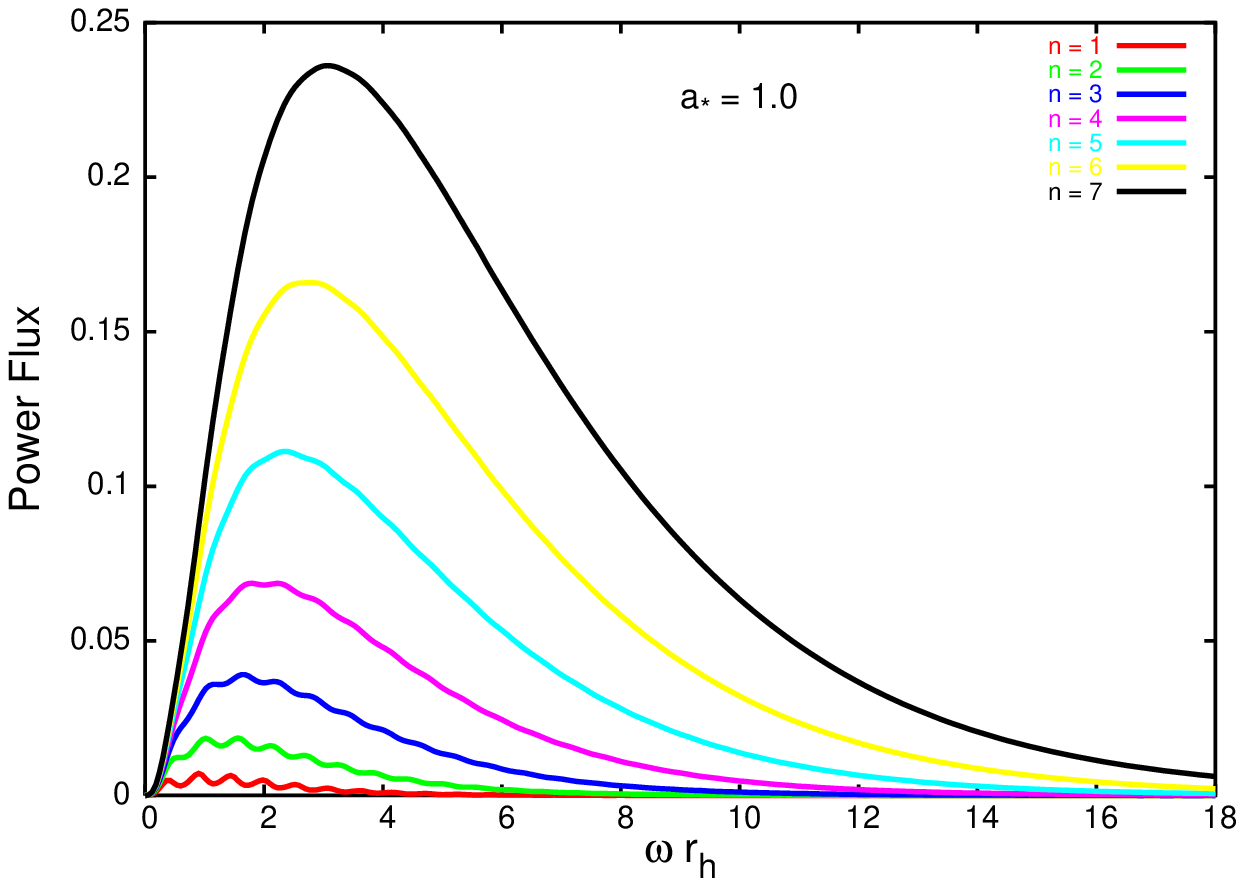}}
\caption{\label{ratesrotplot} Energy emission rates for brane-localised scalar
fields in terms of the angular parameter (left plot) \cite{kanDHKW} and gauge bosons
in terms of the number of extra dimensions (right plot) \cite{kanCKW}.}
\end{center}
\end{figure}

However, unlike the spherically-symmetric case, the gravitational background,
and conse\-quen\-tly the radiation spectra, now depend on two topological
parameters, the angular-mo\-men\-tum parameter $a_*$ of the black hole and the
number of additional spacelike dimensions $n$. We therefore need to explore
the effect that both these parameters have on the radiation spectra. It turns
out that both enhance the energy emission rate: the enhancement factor is of
order ${\cal O}(10)$ in terms of $a_*$ and of order ${\cal O}(100)$ in terms
of $n$. In Fig. \ref{ratesrotplot}, we depict the energy emission rates, for
brane-localised scalars and gauge bosons, in terms of these two parameters,
that clearly exhibit the aforementioned dependence. 

\begin{figure}[b!]
\begin{center}
\mbox{ 
\includegraphics[scale=0.4]{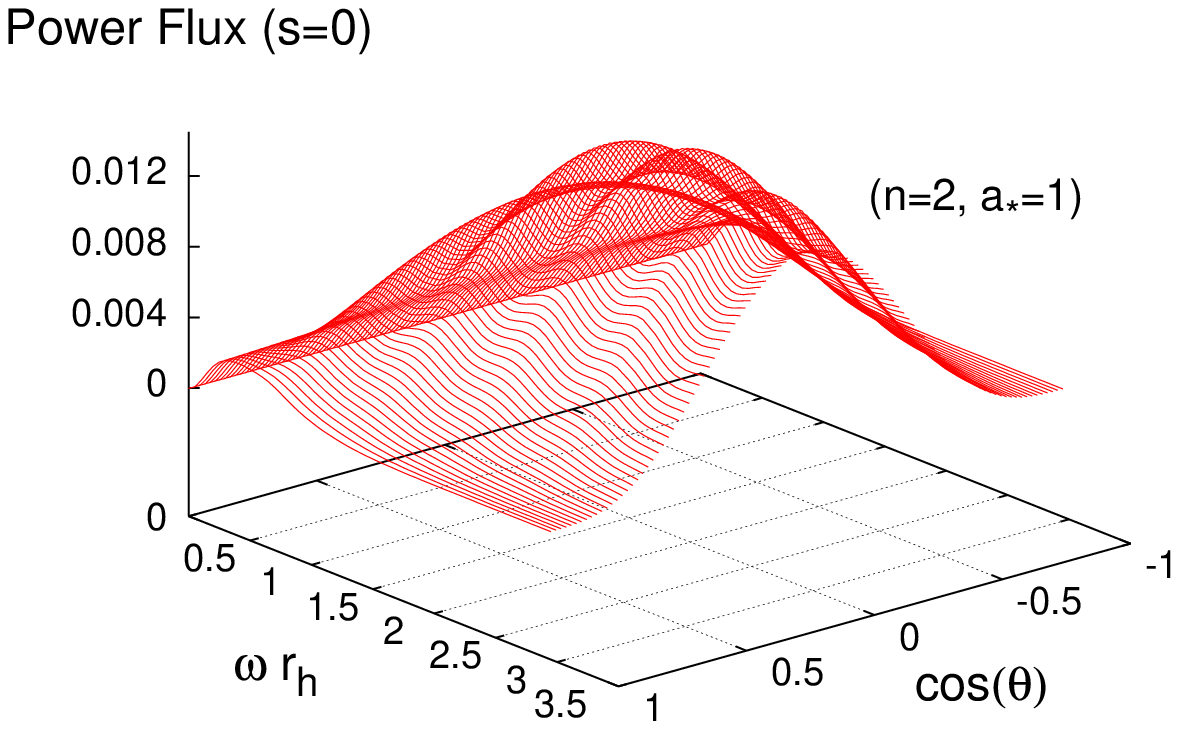}}
{\includegraphics[scale=0.4]{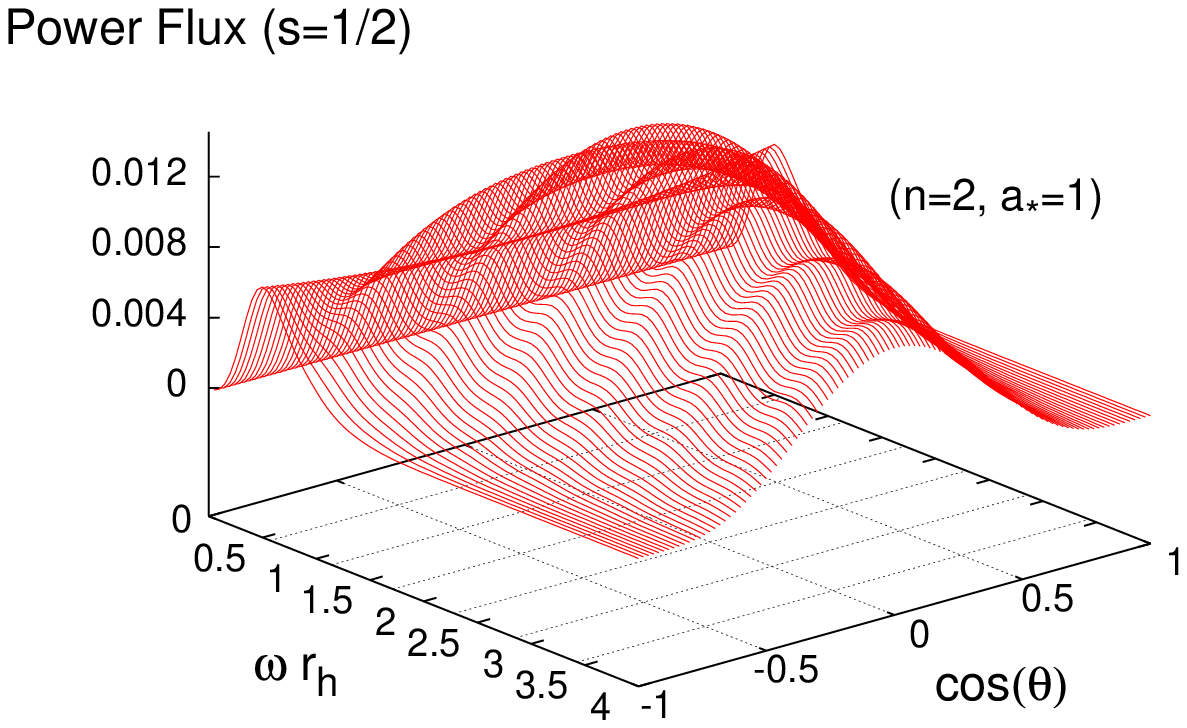}}
{\includegraphics[scale=0.4]{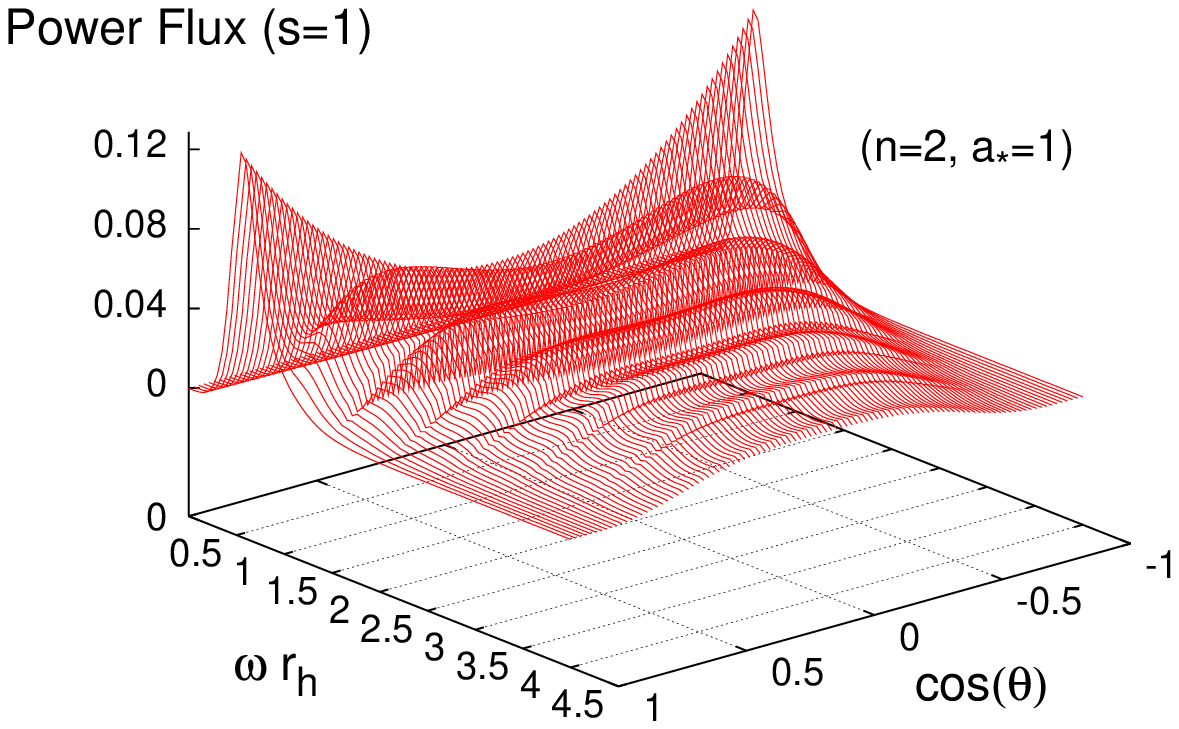}}
\caption{\label{angular-sym-plot} The angular distribution of (a) scalars (upper left plot)
\cite{kanDHKW}, (b) fermions (upper right plot) \cite{kanCKW}, and (c) gauge bosons (lower
plot) \cite{kanCDKW} emitted on the brane by a 6D black hole with $a_*=1$.}
\end{center}
\end{figure}

It is actually the similar effect that the angular momentum of
the black hole and the number of extra spacelike dimensions have on the
radiation spectra that poses the biggest problem in our effort to determine
the value of each. To lift this degeneracy, we need an observable that will depend
rather strongly on only one of them and at the same time be almost insensitive
to the value of the second. The solution is provided by the angular distribution
that characterises the radiation spectra emitted by a rotating black hole.
This information is encoded in the angular equation (\ref{kan-angulareq})
satisfied by the spin-weighted spheroidal harmonics $S^m_{s\ell}$. This equation
was solved numerically \cite{kanCKW,  kanDHKW, kanCDKW} and it was found that
two factors determine the angular distribution: first, the
centrifugal force causes all emitted particles with intermediate and high
frequency to be emitted along the equatorial plane; and second, the
spin-rotation coupling forces the emission of particles with non-vanishing spin
and low frequency to be aligned either in parallel or antiparallel with the
rotation axis (see Fig. \ref{angular-sym-plot}).

\begin{figure}[t!]
\begin{center}
\mbox{ 
\includegraphics[scale=0.55]{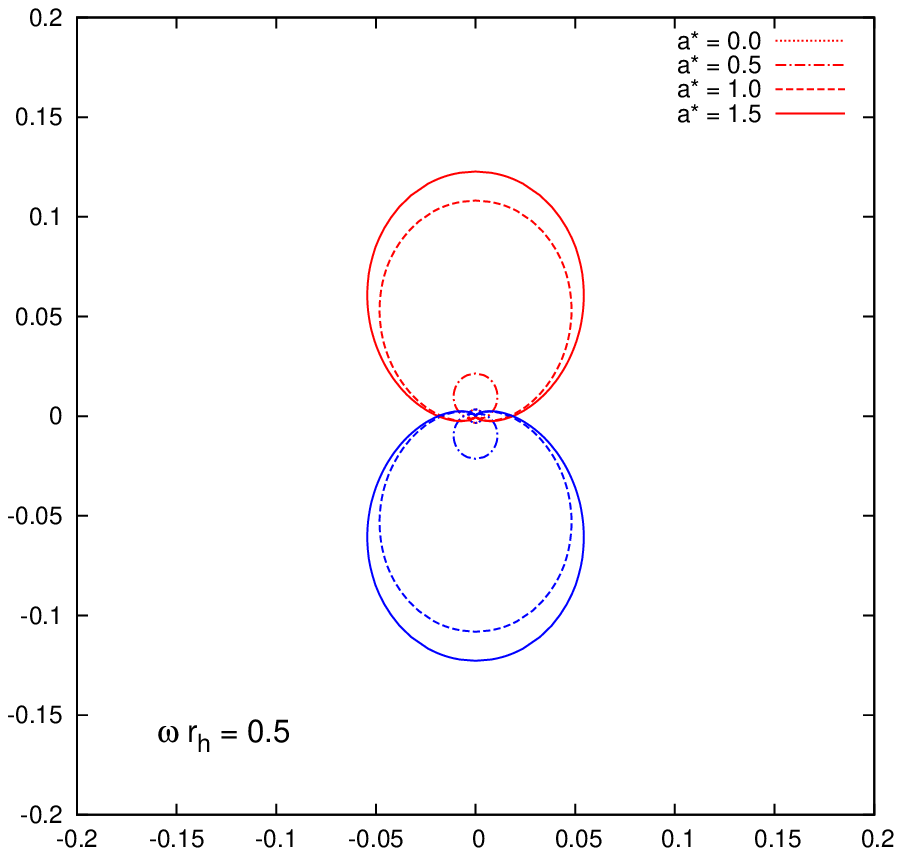}} \hspace*{1cm}
{\includegraphics[scale=0.55]{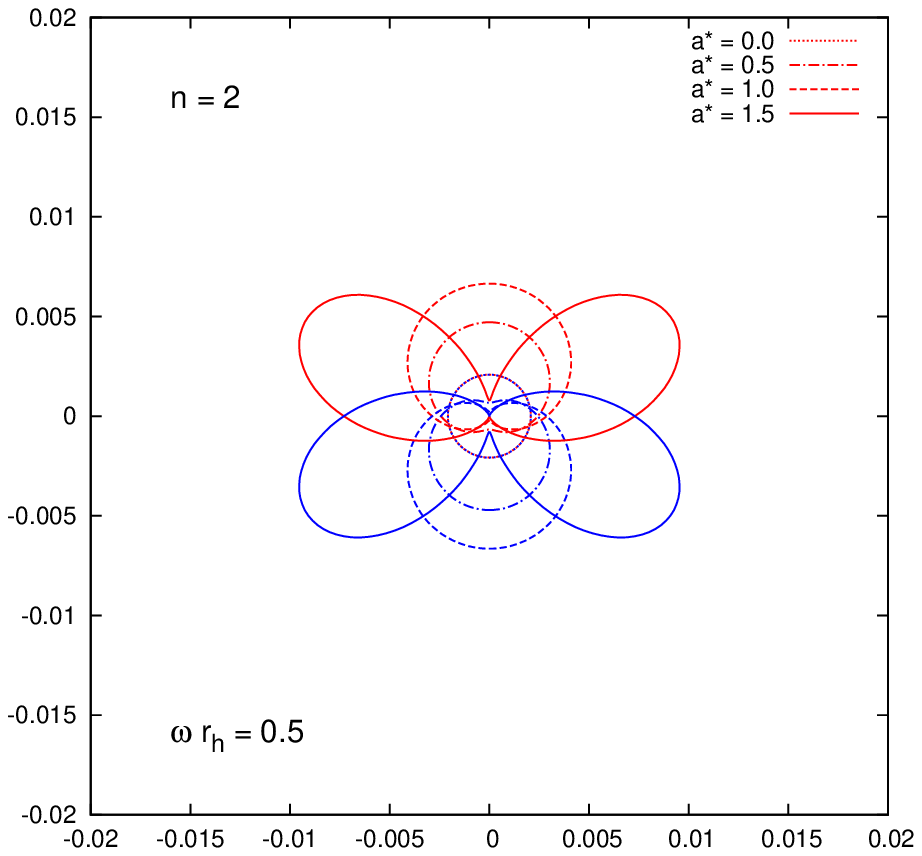}}
\end{center}
\caption{\label{ang-polar} A polar plot with the angular distribution of
(a) gauge bosons (left plot), and (b) fermions (right plot) emitted by a
6-dimensional, rotating black hole on the brane in the energy channel
$\omega r_H=0.5$ \cite{kanCDKW3}. Red and blue curves correspond to positive
and negative helicity particles.}
\end{figure}

The latter effect is more prominent the larger the value of the spin is.
As a result, we expect the alignment of low-energy gauge bosons to be
the best indicator of the orientation of the rotation axis of the black hole
\cite{kanCKW, kanCDKW3}. This is indeed evident from Fig. \ref{ang-polar}(a),
where gauge bosons emitted in the energy channel $\omega r_H=0.5$ are perfectly
aligned along the rotation axis independently of the value of the angular
momentum of the black hole -- note that the rotation axis runs vertically 
along the line $x=0$. Once the orientation of the rotation axis is found,
the angular distribution of low-energy fermions, that is sensitive
to the value of $a_*$  (see Fig. \ref{ang-polar}(b)) \cite{kanCDKW, kanCDKW3,
kanFST}, can now be used in order to determine the value of the angular-momentum
of the black hole itself. As was shown in \cite{kanCDKW3}, the above behaviour
remains unaffected as the number of transverse-to-the-brane extra spacelike
dimensions varies.


\section{Decay of Higher-Dimensional Black Holes in the Bulk}

Another important question is that of the energy balance between the
emission on the brane and that in the bulk. The bulk emission can
never be detected by a brane observer, and any amount of energy emitted
in this channel will be missing energy in a particle experiment.
However, one should know how much energy is spent on the bulk emission as
that determines the amount of energy left for emission on the brane.
According to the assumptions of the brane-world models, Standard-Model
particles are restricted to live on the brane and are therefore emitted,
via the process of Hawking radiation, only on the brane. In the bulk,
we may have only particles that do not carry quantum numbers
under the Standard-Model group, namely gravitons and possibly scalar fields.
We thus need to investigate the emission of both these species of fields
in the bulk during the evaporation process of the black hole.

\begin{figure}[t!]
\begin{center} 
\includegraphics[scale=0.8]{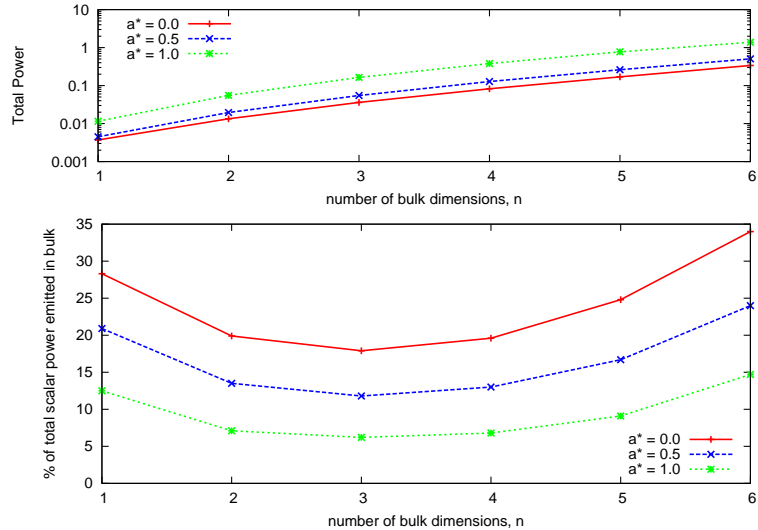}
\caption{\label{kans0abulk} Total power emitted by a rotating black hole in the scalar
channel (upper plot) and the $\%$ of this power emitted in the bulk (lower plot)
\cite{kanCDKW2}.}
\end{center}
\end{figure}

The case of scalar fields propagating in the background of a higher-dimensional
black hole is the easiest one to study. Their equation of motion can be decoupled,
for both the spherically-symmetric and rotating phase, into a radial and an
angular part. The scattering problem may be solved along the same lines as
on the brane and the greybody factors are found both analytically \cite{kanKMR,
kanFS0, kanCEKT4} and numerically \cite{kanHK, kanJP05, kanCDKW2}. For the
spherically-symmetric phase, the relative emissivity, i.e. the ratio of the 
total energy emitted by the black hole per unit time in the bulk over the one
on the brane, is found to depend only on the number of extra dimensions: its value
remains always below unity as $n$ varies from 1 to 7, however, the emission
in the two channels becomes comparable for the highest values of $n$ \cite{kanHK}. 
In \cite{kanCDKW2}, the emission of scalar fields during the spin-down phase was
studied, and it was demonstrated that the total energy output of the black hole
increases with the angular momentum -- however, the relative emissivity reduces
further, compared to the Schwarzschild phase (see Fig. \ref{kans0abulk}), due to
the fact that the enhancement of the greybody factor in the bulk is smaller than
the one on the brane. As a result, the brane dominance persists even in the
spin-down phase, at least in the scalar channel of emission.

To settle this question, one would need to consider also the emission of
gravitons in the bulk. This part of the study is not yet complete: while the 
field equations for gravitational perturbations in a higher-dimensional,
spherically-symmetric background have been derived \cite{kanIKpert}, the
ones in an axially-symmetric spacetime remain largely unknown. For the
Schwarzschild phase, the existing analyses, analytical \cite{kanCNS, kanCEKT1}
as well as numerical \cite{kanCCG, kanJP} ones, have revealed that gravitons carry an
increasingly large amount of energy in the bulk as the number $n$ of the
extra dimensions increases, too. Although the enhancement factor reaches the
value of $10^{6}$, the overall energy carried into the bulk, compared to
the one released in total in the form of SM brane fields, is still sub-dominant. 

\begin{figure}[t!]
\begin{center} 
\includegraphics[scale=0.8]{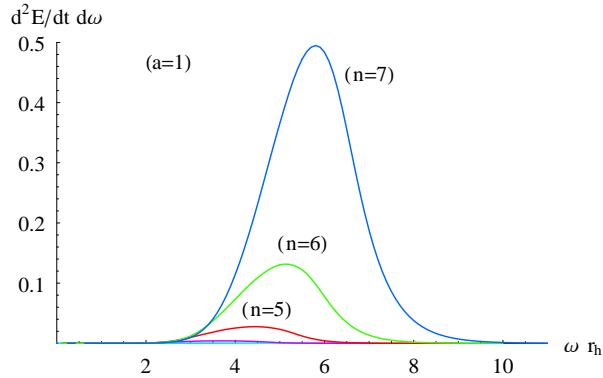}
\caption{\label{kan-grav} Energy emission rate for tensor-type gravitons
in the bulk during the rotating phase \cite{kanKKKPZ}.}
\end{center}
\end{figure}

The above results offer further support to an early work \cite{kanEHM} where
such a claim was made. Nevertheless, since special cases where the bulk channel may
dominate over the brane one are known \cite{kanBGK, kanCCDN}, we still need to
clarify the situation for the gravity emission in the bulk during the rotating
phase. However, the graviton equations are known in only some very particular
cases, one of them being the case of the simply rotating black hole where the
field equations for tensor only gravitational perturbations have been derived
\cite{kanKodama}. In this case, the radiation spectra have the form of
Fig. \ref{kan-grav}, where we depict the tensor-type graviton spectrum in
terms of $n$ for a black hole with $a_*=1$ \cite{kanDCCN, kanKKKPZ}. 
In \cite{kanKKKPZ}, an estimate of the total percentage of energy going into
the bulk tensor graviton channel, compared to the one going into the scalar channel,
was made: for the indicative case of $a_*=1$, the energy carried by gravitons
into the bulk is negligible for low values of $n$, but it reaches the value
of 25\% for $n=5$. In order to draw a conclusive answer on whether the gravitons
may tilt the bulk-to-brane energy balance towards the bulk during the rotating
phase, we clearly need to include in our study the vector and scalar
gravitational perturbations, too.

\begin{figure}[t!]
\begin{center} 
\includegraphics[scale=0.85]{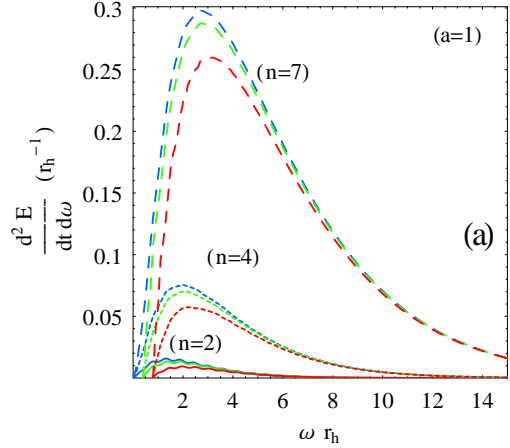}
\caption{\label{kan-massive} Energy emission rates for massive scalar fields
on the brane during the rotating phase for $m_\Phi=0,0.4,0.8$ (from top
to bottom in each set of curves) \cite{kanKP}.}
\end{center}
\end{figure}

Finally, let us report results from a related work where the effect of
the mass of the emitted scalar particles on the energy emission rates both
in the bulk and on the brane, and consequently on the relative emissivity
between the two channels, was studied \cite{kanKP}.
In Fig. \ref{kan-massive}, we depict the indicative cases of the emission of
scalar fields on the brane with $m_\Phi=0,0.4,0.8$ by a black hole with
angular-momentum parameter $a_*=1$ and for three different values of $n$.
As expected, the energy emission rate is reduced as the mass of the emitted
field increases -- clearly, the emission of a massive field demands more energy
to be spent by the black hole, and thus it is less likely to happen. 
However, this suppression is more prominent for the brane channel than for
the bulk one and this gives, at particular cases, a considerable boost to
the bulk-over-brane energy ratio: e.g. for $m_\Phi=0.8$, the bulk-to-brane
relative emissivity can be increased by 34\% if  $a_*=0.5$ and $n=2$ --
as either $a_*$ or $n$ increases, the enhancement takes smaller but still
significant values. 

As is only natural, many experiments looking for physics beyond the Standard
Model have included searches for miniature black holes in their research
programs. Until now, no such effect has been observed with the most recent
limits being the ones derived by the Large Hadron Collider at CERN. Both
CMS \cite{kanCMS} and ATLAS \cite{kanATLAS} collaborations have released data
from proton-proton collisions corresponding to center-of-mass energy of
7 TeV and integrated luminosities of 35 ${\rm pc}^{-1}$ and 36 ${\rm pc}^{-1}$,
respectively. The CMS collaboration saw no excess above the predicted QCD
background, and therefore concluded that, at 95\% CL, no black holes exist
with a minimum mass of (3.5-4.5)\,TeV in models with $n=2,4,6$ and $M_*=(1.5-3.0)$\,TeV.
On the other hand, the ATLAS collaboration, in the absence again of any
events, excluded the existence of black holes in models with $n=6$ and
$M_*=(0.75-3.67)$\,TeV. Nevertheless, as was discussed in section 2, the
window of energy for the creation of black holes might not be open yet
as the minimum required energy for such an effect is around 8 TeV.
Therefore, we should still wait for LHC to reach the ultimate goal of
14 TeV before drawing our final conclusions.

\section{Conclusions}

During the last decade, the topic of the Hawking radiation emission
spectra from higher-dimensional, miniature black holes has undergone
an intensive research. In this talk, I have addressed two questions
that still remain open: the possibility of drawing information on the
fundamental parameters of the higher-dimensional black-hole background,
namely the dimensionality of spacetime in which it was formed and the value
and orientation of angular momentum with which is was formed, and the
determination of the relative bulk-over-brane energy emissivity.
For the former question, we have proposed a solution related to the
angular variation of the emitted radiation that is characteristic of the
emission from a rotating black hole. For the latter question, the
existing results in the literature point to the conclusion that the
brane channel is dominant over the bulk one, during both the spin-down
and the Schwarzschild phase, nevertheless the study of the radiation
spectra for all types of gravitational perturbations is not complete yet.

\begin{acknowledgement} I am grateful to my collaborators
(S. Dolan, M. Casals, R. Konoplya, H. Kodama,
N. Pappas, E. Winstanley and A. Zhidenko) for our enjoyable and fruitful collaborations.
I would also like to thank the organisers of the SEENET-MTP Workshop BW2011 
``Particle Physics from TeV to Planck Scale'' for their kind invitation
to present this talk and for their support.
\end{acknowledgement}

\end{document}